\documentclass[final,1p,times]{elsarticle} 
\usepackage{graphicx} 
\usepackage{amssymb} 
\usepackage{amsthm} 
\usepackage{lineno} 

\journal{Nuclear Physics A} 
\begin{document} 

\begin{frontmatter} 


\title{Energy loss and thermalization of heavy quarks in a strongly-coupled plasma}

\author{C. Marquet$^{a}$, G. Beuf$^{a}$ and B.-W. Xiao$^{b}$}

\address[a]{Institut de Physique Th{\'e}orique, CEA/Saclay, 91191 Gif-sur-Yvette cedex, France}
\address[b]{Nuclear Science Division, Lawrence Berkeley National Laboratory, Berkeley, CA 94720, USA}

\begin{abstract} 

Using the AdS/CFT correspondence, we compute the medium-induced energy loss of a decelerating heavy quark moving through a strongly-coupled supersymmetric Yang Mills plasma. In the regime where the deceleration is small, a perturbative calculation is possible and we obtain the first two corrections to the energy-loss rate of a heavy quark with constant velocity. The thermalization of the heavy quark is also discussed.

\end{abstract} 

\end{frontmatter} 




\section{Motivation}

Nuclear modification factors $R_{AA}$ compare particle production in A+A collisions to a set of independent p+p collisions, such that $R_{AA}$ is unity in the absence of medium effects. In relativistic heavy ion collisions, in which a thermalized quark-gluon plasma (QGP) is formed, the suppressed production of high$-p_T$ particles ($R_{AA}<1$) is interpreted as a consequence of the interaction of the plasma with the hard partons that initiate the hadrons. Indeed, after an energetic parton is produced in the early stages of the collision, it loses energy while propagating through the QGP, and may even thermalize before exiting the medium, meaning less particle production at high-$p_T.$

When one tries to reproduce the value $R_{AA}\simeq 0.2$ observed for light hadrons with perturbative QCD (pQCD) medium-induced energy loss calculations, one obtains that the jet quenching parameter $\hat{q},$ which essentially characterizes how fast energetic partons lose energy, is of order $10\ \mbox{GeV}^2/\mbox{fm}$ \cite{Eskola:2004cr}, a value unnaturally high for a weakly-coupled QGP. Moreover, once this parameter is fixed, pQCD calculations predict that the $R_{AA}$ for heavy hadrons should be higher than it is for light hadrons, since heavy quarks lose energy at a smaller rate than light quarks. The fact that the $R_{AA}$'s for light and heavy hadrons are comparable adds to the puzzle and one wonders whether the pQCD framework is applicable in this context.

This motivated to think about strongly-coupled plasmas. Addressing the strong-coupling dynamics in QCD is an outstanding problem, lattice simulations remain the only method to obtain quantitative results, but are inefficient when real-time dynamics is required, which is the case when analyzing jet quenching. However for a class of non-abelian thermal gauge theories, the AdS/CFT correspondence \cite{Maldacena:1997re,Witten:1998qj,Gubser:1998bc} provides an alternative, and it has become a popular approach to assume that just above the critical temperature $T_c$ where the conformal anomaly in QCD is small, the QGP can be approximated by an $\mathcal{N}=4$ Super Yang Mills (SYM) plasma in the strong coupling regime. Information about real-time dynamics can be obtained within this setup, and in this work we address the energy loss and thermalization problem for heavy quarks propagating through the SYM plasma.

\section{The trailing-string picture}

We consider the large $N_c,$ small gauge coupling $g_{YM}$ limit of the SYM theory:
\begin{equation}
N_c\to\infty\ ,\hspace{0.5cm} g_{YM}\to0\ ,\hspace{0.5cm} \lambda\equiv g_{YM}^2 N_c\mbox{ finite,}
\end{equation}
where the 't Hooft coupling $\lambda$ controls the theory. Then strong coupling means $\lambda>>1.$
In this regime the equivalent string theory in $AdS_5$ space is weakly coupled and weakly curved:
\begin{equation}
g_{YM}\ll1\Leftrightarrow g_s\ll1\mbox{ and }\lambda\gg1\Leftrightarrow R\gg l_s\ ,
\end{equation}
where $g_s$ is the string coupling, $l_s$ is the string length and $R$ is the curvature radius of the AdS space. In this limit of small string coupling and large curvature radius, classical gravity is a good approximation of the string theory, and the background metric corresponding to the thermal SYM theory is
\begin{equation}
ds^2=R^2\left[\frac{du^2}{h(u)}-h(u)dt+u^2(dx^i)^2\right]\equiv G_{\mu\nu}dX^\mu dX^\nu
\end{equation}
with $h(u)=u^2-u_h^4/u^2$ and where $u_h=\pi T$ is a black-hole horizon. The corresponding Hawking temperature $T$ is the temperature of the SYM plasma. The coordinate in the fifth dimension
$u=r/R^2$ has the dimension of momentum and the SYM theory lives on the boundary at
$u=\infty.$

The heavy quark whose energy loss we want to compute is in the fundamental representation, and lives on a brane at $u=u_m=2\pi m/\sqrt{\lambda}\to\infty$ with $m$ the vacuum mass of the quark. A string is attached to it and hangs down towards the horizon. Points on the string can be identified to quantum fluctuations in the heavy quark wave function with virtuality $\sim u.$ Indeed, the quantum dynamics in the SYM theory is mapped onto classical dynamics in the 5th dimension. More precisely, the string dynamics is given by the Nambu-Goto action
\begin{equation}
S=-\frac{\sqrt{\lambda}}{2\pi R^2}\int d\tau d\sigma \sqrt{-g}
\quad\mbox{with}\quad g=\det g_{ab}
\end{equation}
where $\tau$ and $\sigma$ are the worldsheet coordinates and $g_{ab}=G_{\mu\nu}(\partial_a X^\mu)(\partial_b X^\nu)$ is the metric induced on the worldsheet. Let us assume that the quark moves along the $x$ direction and parameterize the space-time
coordinates in the following way: $X^\mu=(t,x,y,z,u)=(\tau,x(\tau,\sigma),0,0,\sigma).$ Then
\begin{equation}
S=-\frac{\sqrt{\lambda}}{2\pi}\int dtdu\sqrt{1-\frac{u^2\dot{x}^2}{h(u)}+u^2h(u)x'^2}
\end{equation}
with $\dot{x}=\partial_t x(t,u)$ and $x'=\partial_u x(t,u).$ The classical equation of
motion $\partial_a\delta{\cal L}/\delta\partial_a X^1=0$ gives 
\begin{equation}
\frac{\partial}{\partial u}\left(\frac{u^2h(u)x'}{\sqrt{-g}}\right)
-\frac{u^2}{h(u)}\frac{\partial}{\partial t}\left(\frac{\dot{x}}{\sqrt{-g}}\right)=0\ .
\label{eqm}\end{equation}
From the solution of this equation, one gets the rate at which the energy flows
down the string:
\begin{equation}
\Phi(t,u)=\frac{\delta{\cal L}}{\delta\partial_\sigma X^0}
=\frac{\sqrt{\lambda}\ R^2}{\sqrt{-g}}u^2h(u)\dot{x}x'\ .\label{elossrate}
\end{equation}

In \cite{Herzog:2006gh,Gubser:2006bz}, the authors imagined using an external force
to compensate the energy loss and pull the quark at a constant velocity $v.$
Writing $x(t,u)=x_0+vt+vF(u),$ they obtained
\begin{equation}
F(u)=\frac{1}{2u_h}\left[\frac{\pi}{2}-\tan^{-1}\left(\frac{u}{u_h}\right)
-\cot^{-1}\left(\frac{u}{u_h}\right)\right]\ .
\end{equation}
In this stationary setup, the corresponding energy-loss rate (\ref{elossrate}) is also independent of $u$ and is therefore identified to the heavy-quark energy loss:
\begin{equation}
-\frac{dE}{dt}=\frac{\sqrt{\lambda}}{2\pi}u_h^2 \gamma v^2\ .
\label{rate}\end{equation}
Investigating this picture in more detail, one finds that the energy loss is due to the 
presence of an induced horizon on the string worldsheet \cite{Chernicoff:2008sa}. This special point in the fifth dimension, located at $u=u_s$ where
\begin{equation}
u_s=\sqrt{\gamma}u_h\quad\mbox{with}\quad\gamma=\frac{1}{\sqrt{1-v^2}}\ ,
\label{satscale}\end{equation}
divides the trailing string into two parts. This is the location where the local speed of light in AdS space coincides with the propagation speed of the string: no light ray emitted from the part of the string below $u_s$ can reach the heavy quark (even if it will reach the boundary at $u_m$), the part of the string with $u<u_s$ is not causally connected to the quark. In \cite{Dominguez:2008vd}, it was shown that the worldsheet horizon corresponds to the saturation scale $Q_s$ on the gauge theory side (the scale at which the scattering of dilute probes off the plasma becomes strong), and that the quantum fluctuations corresponding to points on the string with $u<u_s$ should be thought of emitted radiation, rather than part of the heavy-quark wave function. The picture is identical for a weakly-coupled plasma, but with a different saturation scale \cite{Marquet:2008kr}.

\section{Letting the heavy quark decelerate}

In the following we study the system formed by the upper part of the string, dual to the heavy quark dressed with quantum fluctuations. The energy of this system is
\begin{equation}
E(t)=-\int_{u_s}^{u_m} du\ \frac{\delta{\cal L}}{\delta\partial_\tau X^0}=
\frac{\sqrt{\lambda}R^2}{2\pi} \int_{u_s}^{u_m} du\ \frac{1+u^2 h(u) x^{\prime 2}}{\sqrt{-g}}\ .\label{defE}
\end{equation}
Using the equations of motion one can derive the following energy-conservation equation
\cite{Beuf:2008ep}:
\begin{equation}
\frac{dE}{dt}=\frac{\delta\left. E\right\vert_{du_{s}}}{dt}-\Phi(t,u_s)+\Phi(t,u_m)\ ,
\label{econs}
\end{equation}
where
$\delta\left. E\right\vert_{du_{s}}/dt$ is the energy change of the system due to possible variations of $u_s.$ The energy flow $\Phi(t,u_m)$ represents the work external forces acting on the heavy quark while the energy flow $\Phi(t,u_s)$ represents the energy radiated into the plasma by the dressed quark.

In \cite{Herzog:2006gh}, an external force is acting on the heavy quark and adjusted in order to have a stationary solution with $dE/dt=\delta\left. E\right\vert_{du_{s}}/dt=0.$ The energy flow is also uniform along the string, agreeing with $\Phi(t,u_s)=\Phi(t,u_m):$ the same amount of energy is put into the string at the top than flows at the bottom into the worldsheet horizon, or equivalently is radiatively lost by the heavy quark. In this work we would like to consider the case where the external force is turned off and the heavy quark is allowed to decelerate. Indeed using $\Phi(t,u_m)=0$ in (\ref{econs}) imposes that $-\dot{v}\propto u_h^2/u_m.$ Assuming a slow deceleration, the energy loss for this setup can be computed perturbatively. Although the total energy loss $-dE/dt$ of the upper part of the string includes the contribution from the term
$\delta \left. E\right\vert _{du_{s}},$ we focus on the radiative energy loss $\Phi(t,u_s)$ which is physically more important as it is directly related to the deceleration of the quark, and therefore control how fast thermalization is achieved.

With first-order ($\dot{v}$) and second-order ($\dot{v}^2$ and $\ddot{v}$) corrections, inserting the Ansatz
\begin{equation}
x(t,u)=x_0+\int^t v(t')dt'+v(t)F(u)+\dot{v}(t)\ \zeta (u,v(t))
+\dot{v}^2(t)\ \chi_1\left(u,v(t)\right) +\ddot{v}(t)\ \chi_2\left(u,v(t)\right)
\end{equation}
into the equation of motion (\ref{eqm}), one obtains the first-order deformation of the trailing string:
\begin{equation}
\zeta (u,v)=\frac{\gamma ^{2}}{2u_{s}^{2}}\left[ \tan ^{-1}\left( \frac{u}{%
u_{s}}\right) +\log \left( \frac{u+u_{s}}{\sqrt{u^{2}+u_{s}^{2}}}\right) -%
\frac{\pi }{2}\right] +\frac{F^{2}\left( u\right) }{2}\ .
\end{equation}
This allows to compute the energy flow $\Phi(t,u),$ and imposing $\Phi(t,u_m)=0$ yields
\begin{equation}
\dot{v}(t)=-\frac{v\ u_h^2}{\gamma^2 (u_m-u_s)}
\left[1+ {\cal O}\left(\frac{u_h^2}{(u_m-u_s)^2}\right) \right]\ ,
\end{equation}
and one can now specify our expansion parameter: $u_h/(u_m-u_s).$ In particular our perturbative expansion breaks down if $u_s$ is too close to $u_m,$ meaning if the worldsheet horizon is too close to the flavor brane. Finally the radiative energy loss is
\begin{equation}
\Phi(t,u_s)=\frac{\sqrt{\lambda}}{2\pi}\gamma v^2 u_h^2 \left[1-\frac{(1+\gamma^2) u_h^2 F(u_s)}{\gamma^2 (u_m-u_s)}+ {\cal O}\left(\frac{u_h^2}{(u_m-u_s)^2}\right) \right]\ .
\end{equation}
As $F(u)$ is negative, the first-order correction is positive, meaning that the rate of energy loss gets stronger with time. However this is only efficient for small $v$ because in the ultrarelativistic case $\gamma\gg1$, the first-order correction is suppressed by a factor
$\gamma^{-3/2}.$ As $v$ decreases, the string slowly gets straighter to eventually become that of a static quark but we do not expect our picture to apply that far. Before this happens the heavy quark will thermalize and the discussion should then be modified to include the transverse Brownian motion of the quark. However our calculation allows to estimate how fast thermalization is reached.

Expressions including second-order corrections can be found in \cite{Beuf:2008ep}. For the energy loss, this contribution corresponds to radiation due to the quark deceleration. For $dv/dt,$ one can write:
\begin{equation}
(\mu-\sqrt{\gamma})\frac{dv}{d\tilde{t}}=-v(1-v^2)
\left[1-\frac{\gamma}{(\mu-\sqrt{\gamma})^2}
\left(\frac14+\frac{\pi}8-\frac{\log 2}4\right)\right]\ ,
\label{vdoteq}\end{equation}
where $\tilde{t}=\pi T\ t$ and $\mu=u_m/u_h=2m/(\sqrt{\lambda}T)$ (note that $\mu\gg\sqrt{\gamma}$ in order for our perturbative expansion to be valid). We recover the estimate of the thermalization time made in \cite{Herzog:2006gh,Gubser:2006bz}:
\begin{equation}
\tau=\frac{2m}{\sqrt{\lambda}\pi T^2}\ ,
\end{equation}
but our corrections lead to thermalization faster than what is inferred from the stationary solution.



\end{document}